\documentclass{article}

\usepackage{arxiv}

\usepackage[utf8]{inputenc} 
\usepackage[T1]{fontenc}    
\usepackage{hyperref}       
\usepackage{url}            
\usepackage{booktabs}       
\usepackage{amsfonts}       
\usepackage{nicefrac}       
\usepackage{microtype}      
\usepackage{amsmath,amssymb,amsfonts}
\usepackage{cleveref}       
\usepackage{lipsum}         
\usepackage{cite}

\usepackage{algorithmic}
\usepackage{subcaption}
\usepackage{graphicx}
\usepackage{url}
\usepackage{textcomp}
\usepackage{comment}
\usepackage{xcolor}
\usepackage{balance}

\title{Hiding in Plain Sight: Reframing Hardware Trojan Benchmarking as a Hide\&Seek Modification}

\date{}

\author{ \href{https://orcid.org/my-orcid?orcid=0000-0002-0134-8418}{\includegraphics[scale=0.06]{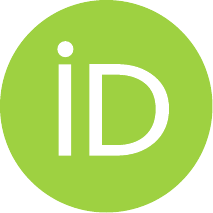}\hspace{1mm}Amin ~Sarihi}, \\
	Klipsch School of Electrical\\ and Computer Engineering\\
	New Mexico State University\\
	\texttt{sarihi@nmsu.edu} \\
	\And
	\href{https://orcid.org/0000-0003-2647-2797}{\includegraphics[scale=0.06]{orcid.pdf}\hspace{1mm}Ahmad ~Patooghy} \\
	Computer Systems Technology\\
	North Carolina A\&T  University\\
	\texttt{apatooghy@ncat.edu}\\
	\And
	\href{https://orcid.org/0000-0002-3741-0201}{\includegraphics[scale=0.06]{orcid.pdf}\hspace{1mm}Peter ~Jamieson} \\
	Department of Electrical\\ and Computer Engineering\\
	Miami University\\
	\texttt{jamiespa@miamioh.edu}\\
	\And
	\href{https://orcid.org/0000-0001-8027-1449}{\includegraphics[scale=0.06]{orcid.pdf}\hspace{1mm}Abdel-Hameed A. ~Badawy} \\
	Klipsch School of Electrical\\ and Computer Engineering\\
	New Mexico State University\\
	\texttt{badawy@nmsu.edu} \\}

\renewcommand{\shorttitle}{Hardware Trojan Insertion Using Reinforcement Learning}

\begin{document}
\maketitle

\begin{abstract}
	This work focuses on advancing security research in the hardware design space by formally defining the realistic problem of Hardware Trojan (HT) detection. The goal is to model HT detection more closely to the real world, \textit{i.e.}, describing the problem as The Seeker's Dilemma where a detecting agent is unaware of whether circuits are infected by HTs or not. Using this theoretical problem formulation, we create a benchmark that consists of a mixture of HT-free and HT-infected restructured circuits while preserving their original functionalities. The restructured circuits are randomly infected by HTs, causing a situation where the defender is uncertain if a circuit is infected or not. We believe that our innovative benchmark and methodology of creating benchmarks will help the community judge the detection quality of different methods by comparing their success rates in circuit classification. We use our developed benchmark to evaluate three state-of-the-art HT detection tools to show baseline results for this approach. We use Principal Component Analysis to assess the strength of our benchmark, where we observe that some restructured HT-infected circuits are mapped closely to HT-free circuits, leading to significant label misclassification by detectors.
\end{abstract}


\section{Introduction}
 Hardware Trojans (HTs) are a threat to digital electronics in general and the design and manufacturing of Integrated Circuits (ICs), in particular\cite{shakya2017benchmarking}. HTs are unwanted modifications in the design or manufacturing of an IC such that the chip’s expected behavior is altered.  Potential impacts of security breaches through HTs have pushed researchers to look into the methods and algorithms to detect HTs in ICs in the early stages of the design and manufacturing~\cite{ hasegawa2017trojan, pan2021automated, gohil2022deterrent,yu2021hw2vec,sarihi2023multi}. Despite the novelty, the field still needs a formal definition of the problem that mirrors the real-world problem of HT detection.

\begin{figure*}[!ht]
    \centering
    
    \includegraphics[scale=0.51]{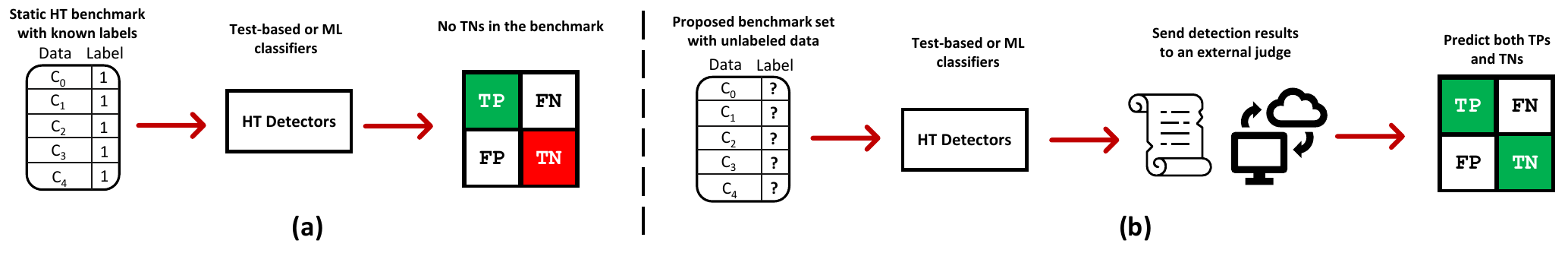}
    \vspace{-2mm}
    \caption{Comparison of a) current HT detection approaches with static HT benchmarks that only contain HT-infected circuits vs. b) our proposed HT detection flow, including restructured benchmarks with and without HTs}
    \label{fig:benchmark}
\end{figure*}

In this paper, our goal is to advance the state-of-the-art in HT detection by defining the problem of HT insertion/detection in digital ICs as a game between two players. Our problem statement is rooted in the \textit{Hide\&Seek} problem on a graph. We call this new formulation “The Seeker’s Dilemma” as it more closely resembles the problem of HT detection from the perspective of real-world manufacturers and distributors of ICs. We take “The Seeker’s Dilemma” approach in creating realistic HT benchmarks to address the shortcomings of existing HT benchmarks. In current benchmarks, the HT location and size are known a priori to researchers such as~\cite{trusthub,cruz2018automated,gohil2022attrition}. This enables the defense side to fine-tune their HT detectors to showcase strong HT detection rates. We believe that a standard benchmark should contain several HT-infected and HT-free instances to provide better dataset balance for researchers. Figure~\ref{fig:benchmark} shows the difference between existing benchmark approaches and our proposed method.

In this work, we assume that the attacker relies on Third-Party Electronic Design Automation (3P-EDA) tools to insert HTs at the synthesis stage\cite{xue2020ten}. The EDA tool enables the attacker to restructure the circuit’s netlist while maintaining the circuit functionality unchanged. It is worth mentioning that functional restructuring used in this study is inherently different from hardware obfuscation in the logic locking context~\cite{hoque2020hardware}. We utilize ABC~\cite{brayton2010abc}, an open-source logic optimization tool, leveraging its various representations and optimization techniques to modify the structure of circuits. Our released benchmark will be called  \textit{Seeker1}, and it will be publicly available for researchers to test their tools~\cite{github}. Our methodology is described so that we can employ this benchmark creation approach to more realistic circuits such as microcontrollers and microprocessors.

The contributions of this work are:
    1) We introduce The Seeker’s Dilemma, a formal definition of HT detection similar to finding HTs in real-world scenarios.
    2) We introduce \textit{Seeker1} (a benchmark with potentially hidden HTs), and the methodology we use to create this benchmark to help improve HT detection.
    3) We use Principal Component Analysis (PCA) to analyze the significance of \textit{Seeker1} from a Machine Learning (ML) training perspective.

\section{Related Work}
\label{sec:related}
HT detection methods can be categorized based on their dependence on a golden model. If security engineers have access to a golden model or its specification, they will look for any deviations in the functionality of the devices from the expected standard behavior when applying test vectors.  Previous work such as ~\cite{pan2021automated,sarihi2023multi,sarihi2024trojan,gohil2022deterrent} employ analytical and ML-based approaches to generate test vectors that they believe would best expose the existence of malicious HTs. When interpreting HT classification results, we encounter 4 possible outcomes: \textit{FP} (false positive), \textit{FN} (false negative), \textit{TP} (true positive), \textit{TN} (true negative). Test-based HT detection would never lead a security engineer to a FP (false positive) case. In addition to using test vectors, other ML-based approaches have also been used extensively towards HT detection when golden models are not available~\cite{salmani2016cotdpaper,hasegawa2017trojan,yu2021hw2vec}. The accuracy of such approaches solely rely on the data on which they were trained (which is the Trust-hub\cite{trusthub} benchmark in most cases). These detectors can result in all four classification outcomes.

HT benchmarks have also been the subject of study by various researchers. Trusthub~\cite{salmani2013design,shakya2017benchmarking} was among the first ones where several HT designs are available to study. Despite its valuable contribution to the HT research community, the benchmark lacks the size and variety needed to push the detection field forward~\cite{cruz2018automated}. Other researchers have made efforts to address Trusthub's shortcomings by developing automated tools that generate HT benchmarks~\cite{cruz2018automated,sarihi2022hardware,gohil2022attrition,sarihi2024trojan}. While pushing the HT field forward, the newly introduced benchmarks are heavily unbalanced where the number of HT-infected data points outnumber the HT-free ones, which leads to training biased ML-based HT detectors. Our work strives to cover the to deficiencies of the previous literature by introducing \textit{Seeker1}. The details of our benchmark will be explained in the rest of the paper. 
\vspace{-1mm}
\section{The Seeker’s Dilemma}
\vspace{-1mm}
\label{sec:seek}
The \textit{Hide\&Seek} problem as related to cybersecurity \cite{chapman2014playing} in the HT domain (a situation which we call ``The Seeker’s Dilemma'') is that given an IC or netlist, the \textit{Seeker} ($S$), does not know whether an HT has been hidden in the netlist in a two-player game where $H$ is the \textit{Hider} and $S$ the \textit{Seeker} (detector). We define $k = |\mathcal{H}|$ as the number of objects hidden, and in the Seeker’s Dilemma, $k$ hidden objects have the condition \underline{$k \geq 0$}. Moreover, \underline{the value of $k$ is unknown by the \textit{Seeker}}. Adding this condition and the unknown information transforms the problem into what we define as ``The Seeker’s Dilemma'', and from a complexity theory perspective, makes the problem significantly harder from a real-world perspective. The Seeker’s Dilemma has very different strategical implications such that the \textit{Hider} hides $k$ objects (could be nothing if $k=0$) and the \textit{Seeker} ($S$) searches for hidden objects in $L$ queries or moves, where $H$ tries to maximize $L$ and, conversely, $S$, tries to minimize $L$.

The major challenge with existing HT benchmarks is that these circuits are known-knowns in terms of the existence of HTs, \textit{i.e.}, almost in all cases, $k = 1$. This means that both the inserting and detecting sides know the situation, which we believe oversimplifies the problem. Despite being very helpful to the community, Trusthub benchmarks~\cite{trusthub} fall into this class. To address this gap, we have created The Seeker’s Dilemma HT benchmark (we call it \textit{Seeker1}) featuring $k \geq 0$. The \textit{Seeker1} benchmark suite will be publicly available.
\subsection{Benchmark Generation}

We have selected the original circuits for \textit{Seeker1} from ISCAS-85’s~\cite {hansen1999unveiling} combinational designs that have been used throughout CAD research community~\cite{cruz2018automated,gohil2022attrition,gohil2022deterrent,yu2021hw2vec}. This benchmark has been widely used to evaluate the effectiveness of logic synthesis tools, technology mapping algorithms, test generation algorithms, timing analysis, various optimization techniques for digital circuits, and HT detection. So, we believe that ISCAS-85 is the first candidate to generate \textit{Seeker1}. 

\begin{figure*}
    \centering
    \begin{subfigure}{0.48\textwidth}
        \centering
        \includegraphics[width=\linewidth]{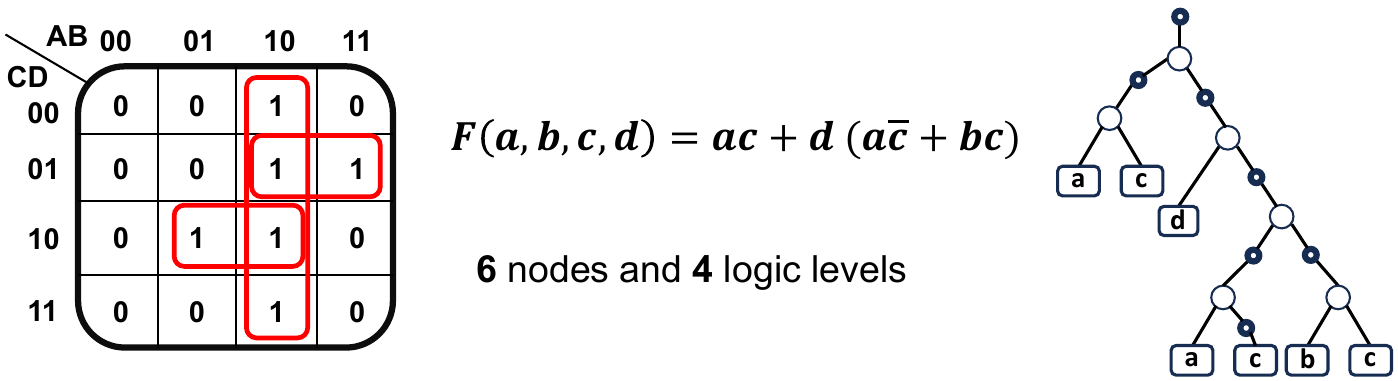}
        \caption{AIG representation \#1}
        \label{fig:subfig_AIG_1}
    \end{subfigure}
    \hfill
    \begin{subfigure}{0.48\textwidth}
        \centering
        \includegraphics[width=\linewidth]{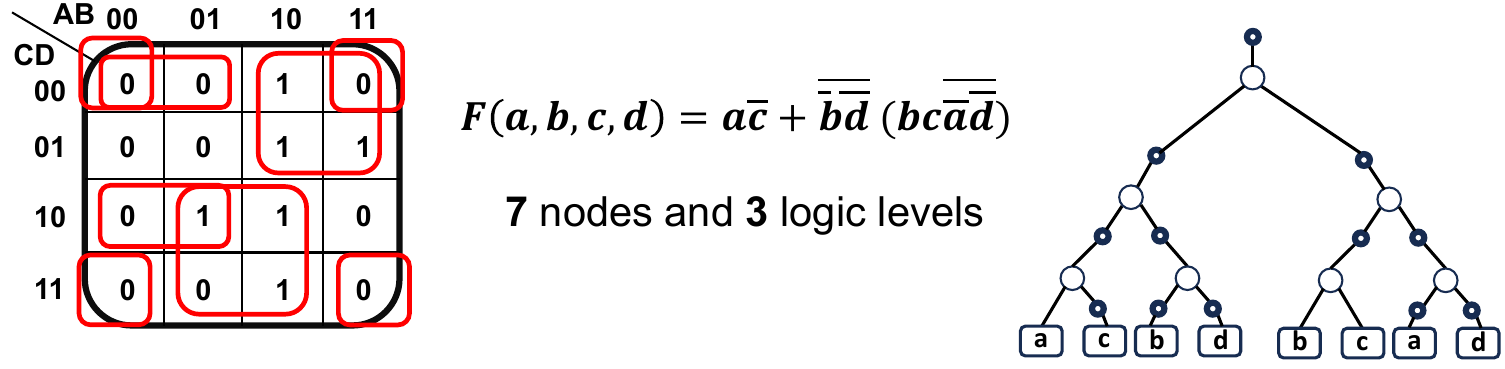}
        \caption{AIG representation \#2}
        \label{fig:subfig_AIG_2}
    \end{subfigure}
    \caption{Two representations of a circuit with the same truth table. Representation \#1 is aimed at improving the area in terms of comparatively fewer nodes, while representation \#2 enhances delay with fewer logic levels}
    \label{fig:AIG}
    \vspace{-6mm}
\end{figure*}

The AND-Inverter Graph (AIG) format is used to facilitate the functional restructuring of our target circuits. AIG represents circuits with two-input \textit{AND} gates and \textit{NOT} gates~\cite{mishchenko2018integrating}, derived using DeMorgan’s rule. AIGs, though non-canonical, are used in algorithms to optimize area, delay, and formal equivalence checking~\cite{chowdhury2021openabc}. An example is shown in Figure~\ref{fig:AIG}. We use ABC~\cite{brayton2010abc} and employ $18$ functional restructuring methods to alter ISCAS-85 circuits, hiding HTs and complicating detection. Our benchmark includes $8$ ISCAS-85 circuits ($c880$, $c1355$, $c1908$, $c2670$, $c3540$, $c5315$, $c6288$, $c7552$) with variably added HTs, details of which are withheld to prevent reverse-engineering. We use the HT circuits generated by an RL framework explained in~\cite{sarihi2022hardware,sarihi2024trojan}. We pick 100 inserted HTs from each circuit and create 18 versions of each. We also use ABC to functionally restructure HT-free circuits to add further complexity to \textit{Seeker1}. 
\vspace{-3mm}
\section{Analysis of our Benchmark}
\label{R}

To analyze our first Seeker’s Dilemma benchmark, \textit{Seeker1}, we test the benchmark against existing HT detection tools. We use three different HT detection strategies/tools to measure the quality of inserted HTs: the test vectors from RL\_HT\_DETECT developed in~\cite{sarihi2023multi,sarihi2024trojan}, the test vectors from DETERRENT proposed in~\cite{gohil2022deterrent}, and the open-source HW2VEC~\cite{yu2021hw2vec}. RL\_HT\_DETECT and DETERRENT are test-based HT detectors requiring a golden model, which is resilient against functional restructuring techniques as these do not alter circuit functionality. HW2VEC, a GNN-based HT detector, extracts behavioral features from hardware designs to train a binary classifier on a dataset with $200$ features, with its performance depending on the dataset quality and quantity. Next, we train a binary classifier with the following training data:
1) The \textit{TJ\_RTL} dataset used in~\cite{yu2021hw2vec} for training HW2VEC. This dataset contains communication protocols and encryption algorithms from Trusthub. The dataset contains $\textbf{26}$ HT-infected and $\textbf{11}$ HT-free instances. The rest of the paper refers to the HT detection reports under this scenario as \textit{$S_1$}.
2) 
We add two versions of ISCAS-85 HT-free benchmarks~\cite{cruz2018automated,hansen1999unveiling} to \textit{TJ\_RTL} to make the dataset labels more balanced. This action adds $\textbf{16}$ more HT-free instances to the previous set. We refer to the HT detection reports under this scenario as \textit{$S_2$} hereafter.

\begin{figure}[!t]
\centering
\includegraphics[width=.75\linewidth]{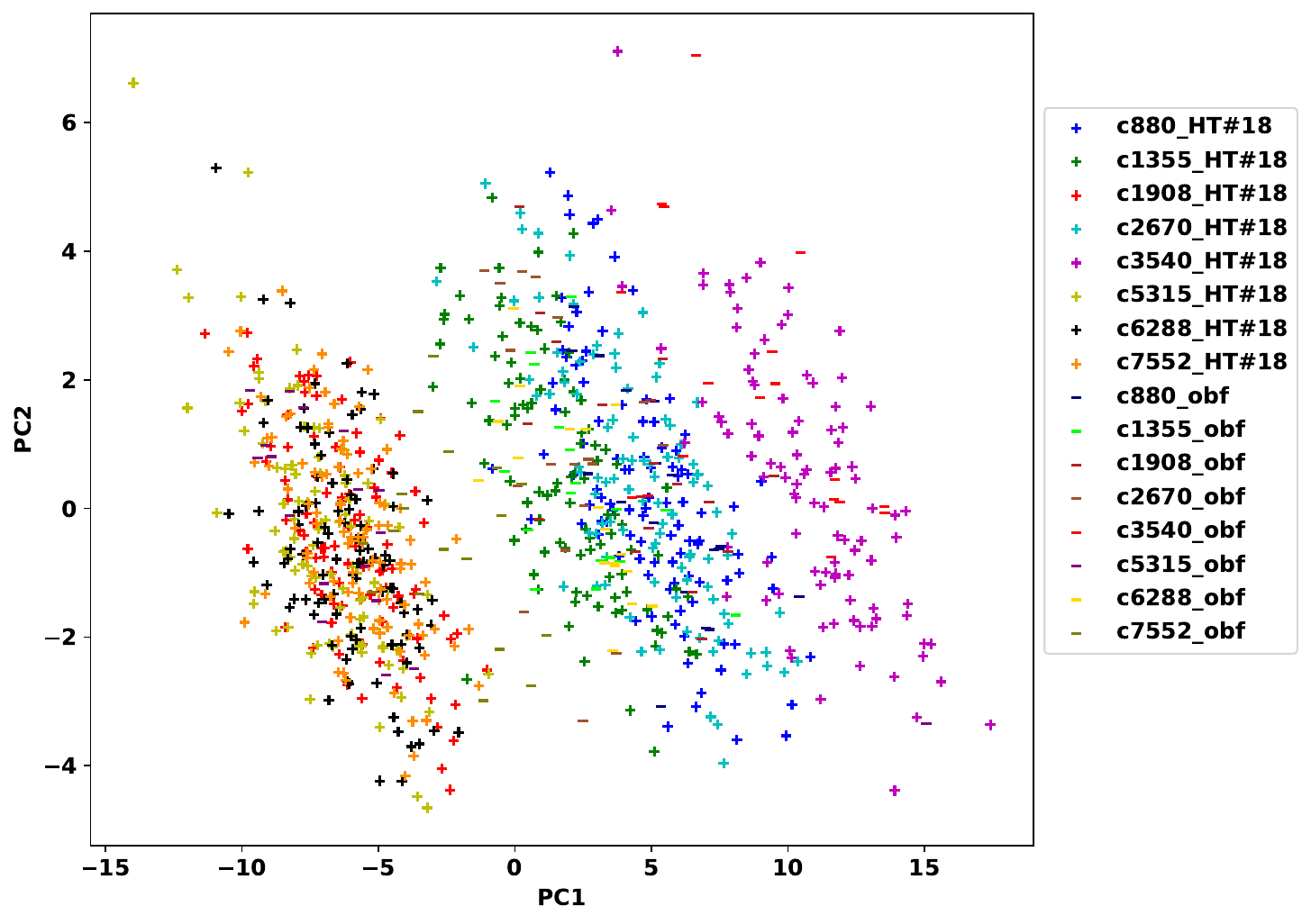}
\caption{PCA analysis of hidden HTs vs.\ clean functionally transformed ISCAS-85 circuits}
\label{fig:ABC_HT_vs_ABC_clean}
\end{figure}

Figure~\ref{fig:ABC_HT_vs_ABC_clean} shows the outcome of the PCA analysis~\cite{bro2014principal} for the functionally restructured versions of both infected (ending with \_HT\#18 suffix) and clean (ending with \_obf suffix) ISCAS-85 circuits with restructuring technique number 18. We transform each circuit with HW2VEC and extract $200$ attributes for each one. PCA is used to reduce dimensionality to a handful of principal components that collectively explain a significant portion of the total variance. As can be seen, the HT-free instances (marked as -) are distributed among the HT-infected circuits (marked as +) when we plot PC1 against PC2. It is hard to find a boundary upon which these two classes can be separated. To further study this, we train HW2VEC with $S_1$ and $S_2$, and investigate the classification accuracy of HT-free data points. Among the ISCAS-85 benchmark circuits, the restructured HT-free instances of $c5315$, $c6288$, and $c7552$ have the highest \textit{FP} rates of 80\%, 40\%, and 45\% on average, while the respective figures for the remaining circuits are under 20\%. This experiment emphasizes the need for a diversified dataset for training HT detectors.

\begin{figure*}[t!]
\centering
    
\begin{subfigure}[t]{0.49\textwidth}
    \centering
    \includegraphics[width=\linewidth]{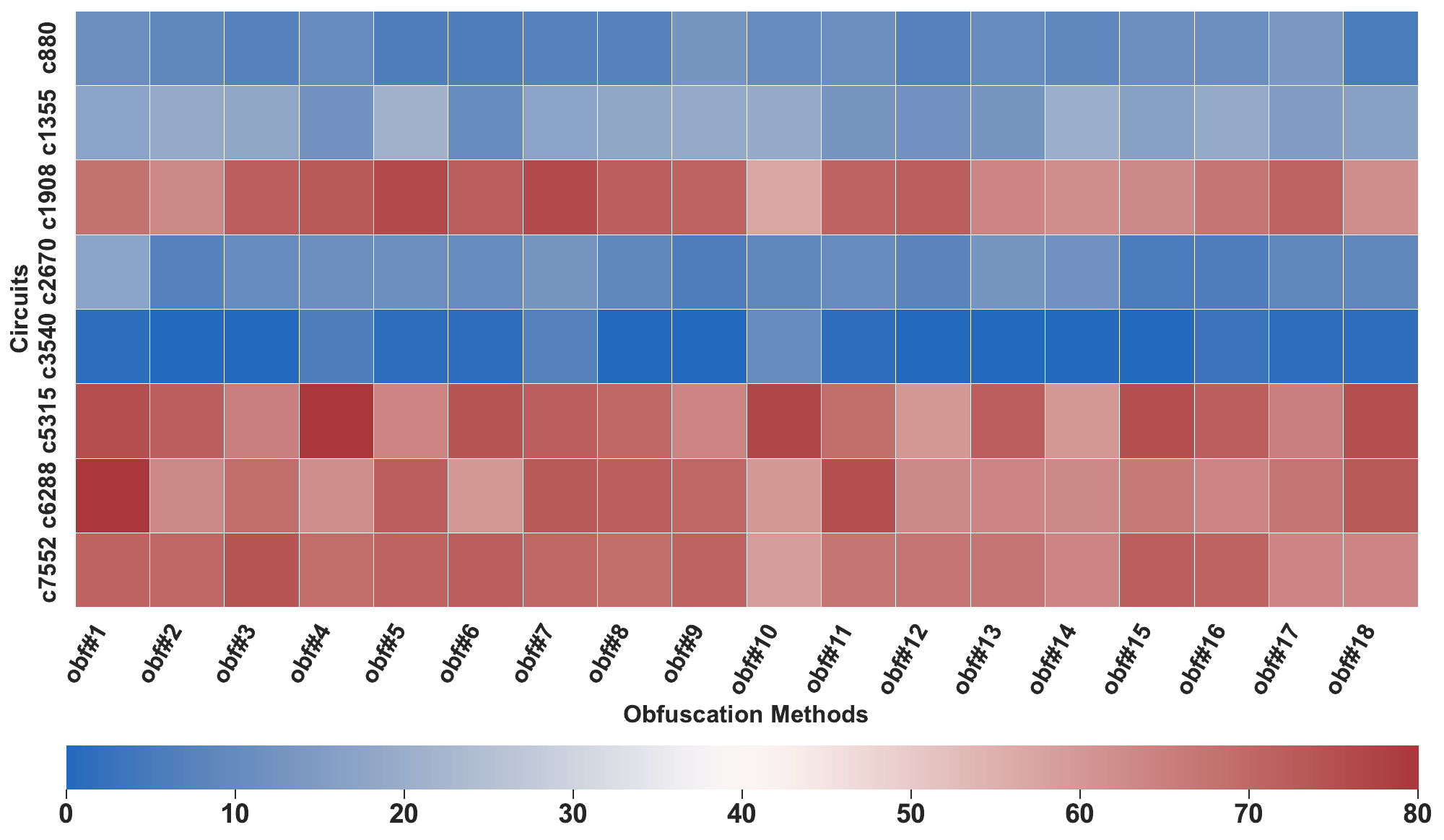}
    \caption{The HT detection accuracy spans between 0\% and 80\% for $S_{1}$}
    \label{fig:heatmap_TJ_RTL}
\end{subfigure}
\vspace{1mm}
\begin{subfigure}[t]{0.49\textwidth}
    \centering
    \includegraphics[width=\linewidth]{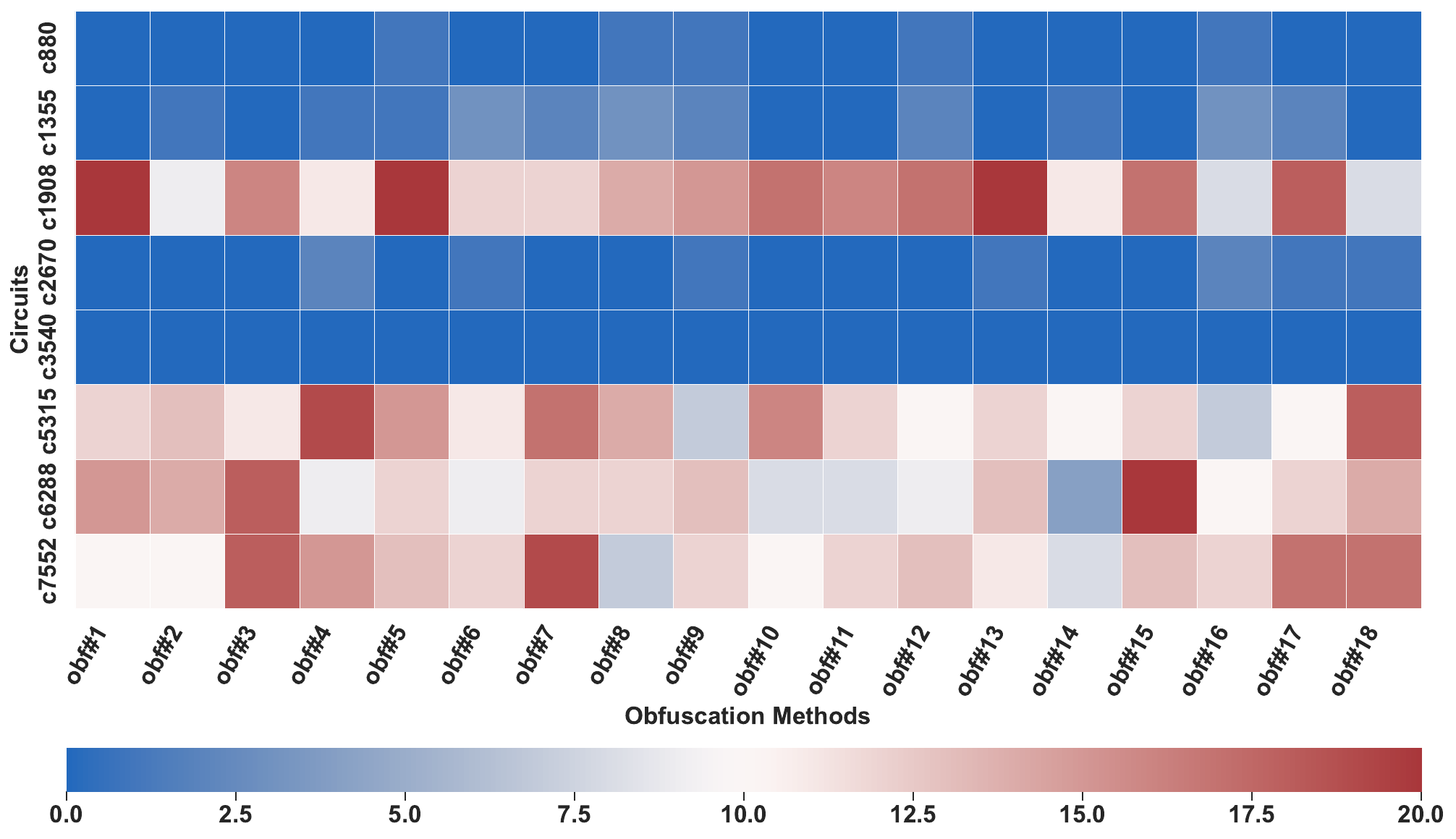}
    \caption{The HT detection accuracy spans between 0\% and 20\% for $S_{2}$}
    \label{fig:heatmap_TJ_RTL_ISCAS}
\end{subfigure}
\vspace{-3mm}
\caption{Detection accuracy of 18 functionally equivalent transformation methods trained with a) $S_{1}$ and b) $S_{2}$ for ISCAS-85.}
\label{fig:HT_detection_heatmap}
\end{figure*}

Figure~\ref{fig:HT_detection_heatmap} shows the heatmap of HT detection accuracy percentages ($TPs$) for the functionally restructured HT-infected circuits using HW2VEC trained with both $S_{1}$ and $S_{2}$. Each circuit-functional-restructuring method pair contains $100$ HT-infected circuits generated by the RL-inserter and functionally equivalent transformations using ABC. The detection accuracy in Figure~\ref{fig:heatmap_TJ_RTL} ranges between $0\%$ and $80\%$ for $S_{1}$ while the same figure ranges between $0\%$ and $20\%$ in Figure~\ref{fig:heatmap_TJ_RTL_ISCAS} for $S_{2}$. In both detection scenarios, the circuits are divided into two groups: 
\{$c880$, $c1355$, $c2670$, and $c3540$\}, \{
$c1908$, $c5315$, $c6288$, and $c7552$\}.

While HW2VEC detects up to $80\%$ of HTs in the second group under $S1$, it significantly underperforms with the first group. The situation worsens under $S2$, where the detector fails to classify HT-infected circuits in group `1` while the figures are only slightly better for group ‘2’. The underlying reason can be sought with the mixture of labels in each $S_{1}$ and $S_{2}$ and the unseen hidden data. The extra HT-free labels in $S_{2}$ bias the detector to classify more instances as HT-free.



We also compare \textit{Seeker1} with two existing benchmarks introduced in~\cite{cruz2018automated,gohil2022attrition}. Both benchmarks only contain HT-infected circuits and HTs are inserted using low signal switching nets. For~\cite{cruz2018automated}, we train HW2VEC under $S_{1}$ and we report the detection accuracies for 4 reported ISCAS-85 benchmarks, \textit{c2670, c3540, c5315, c6288}. The detection figures are 100\%, 0\%, 70\%, and 0\%, respectively. In~\cite{gohil2022attrition}, there are two ISCAS-85 benchmarks: $c6288$ and $c7552$. The detection rates are 10\% and 90\%, respectively. Compared with Figure~\ref{fig:heatmap_TJ_RTL}, \textit{Seeker1} evades detection more consistently throughout the entire benchmark. It is important to note that the insertion criteria of~\cite{sarihi2024trojan,sarihi2022hardware} are inherently different from that of ~\cite{cruz2018automated,gohil2022attrition}.  In the future, we plan to investigate the impact of various HT insertion and functional restructuring strategies on HT detectors.

\begin{figure}[!t]
\centering
\includegraphics[width=.75\linewidth]{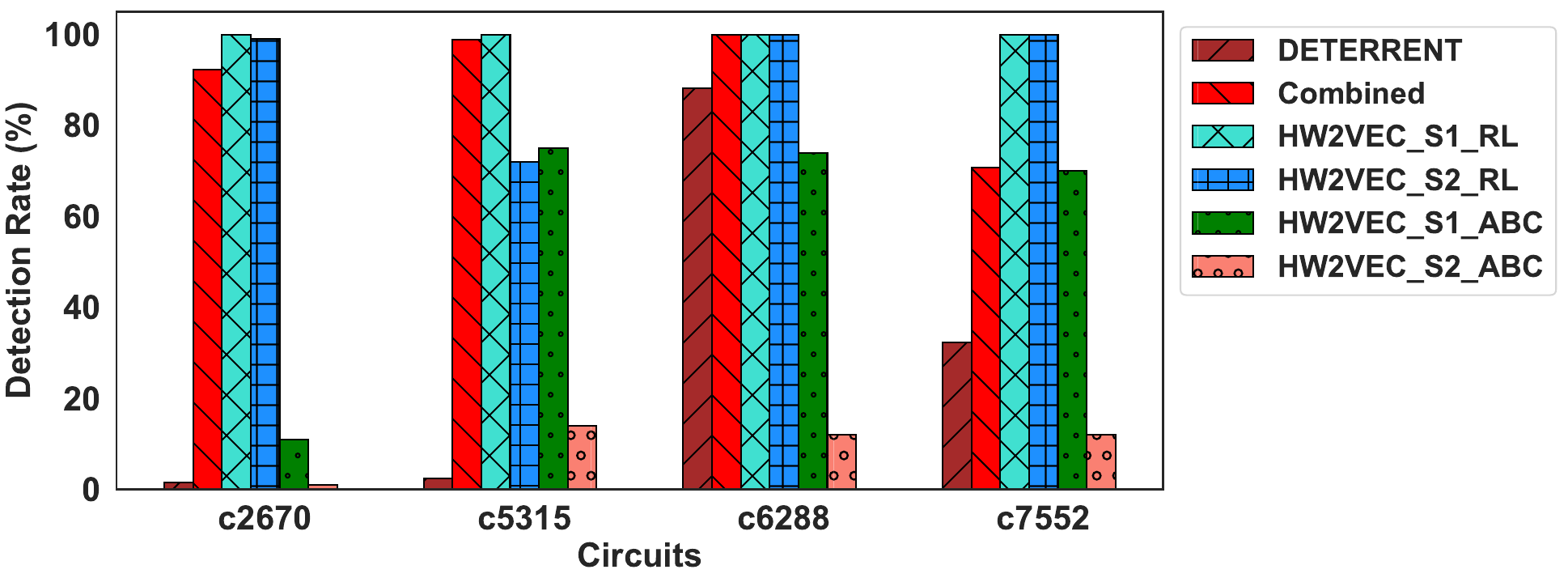}
\caption{The detection rate (TPs) of DETERRENT~\cite{gohil2022deterrent}, RL\_HT\_DETECT~\cite{sarihi2024trojan}, and HW2VEC~\cite{yu2021hw2vec} for hidden HT-infected ISCAS-85 circuits}
\label{fig:joint_detection}
\end{figure}

Figure~\ref{fig:joint_detection} shows the detection percentage for RL\_HT\_DETECT (\textit{Combined}), DETERRENT, and HW2VEC for the original HT-infected circuits (ending with \_RL suffix) and their ABC-functionally equivalent transformed versions (ending with \_ABC suffix). The $x$ axis shows the benchmark circuits and the $y$ axis shows the HT detection accuracy (\textit{TPs}) as a percentage. To fairly compare against DETERRENT, we only mention the four ISCAS-85 benchmarks studied in the DETERRENT paper. 
As can be seen, the detection accuracy of RL\_HT\_DETECT is higher than DETERRENT in all four circuits; however, the difference is more substantial in $c2670$ and $c5315$. The reason can be sought into  multi-criteria~\cite{sarihi2023multi} versus single criterion~\cite{gohil2022deterrent} HT detection. As for HW2VEC's detection rate under $S_{1}$ and $S_{2}$ for the baseline RL benchmarks is nearly $100\%$. The situation differs for the functionally equivalent transformed HTs, with lower HT detection rates. 
 
\vspace{-3mm}

\section{Conclusion}
\label{C}
This work defines HT detection as a Hide\&Seek game termed The Seeker’s Dilemma, highlighting the challenge of detecting infected circuits without prior knowledge. Using this paradigm, a new benchmarking strategy is proposed for evaluating HT detection methods more accurately. The innovative problem statement and strategy aim to foster new ideas and improve quality assessments in HT detection. A combinational benchmark was created and released to guide future work in HT detection and insertion. Existing HT detection methods were tested on this benchmark, marking the first such evaluation. Future plans include training a more robust HT detector against data variations.
\vspace{.4mm}
\section*{Acknowledgments}
This work has been partially funded by NSF grants 2219680 and 2219679.

\bibliographystyle{ACM-Reference-Format}
\bibliography{references}

\end{document}